\documentclass[reprint,aip,pop,amsmath,amsfonts,amssymb,floatfix]{revtex4-1}

\usepackage{graphicx,xcolor,latexsym}

\newcommand{\MyPaperTitle}{Molecular Dynamics Simulations of Field Emission From a Prolate Spheroidal Tip}

\usepackage[colorlinks=true, linkcolor=blue, citecolor=blue, urlcolor=blue, bookmarks=true, breaklinks=true,
            pdftitle={\MyPaperTitle},
            pdfauthor={Kristinn Torfason}]{hyperref}
\usepackage[all]{hypcap}
\usepackage{bookmark}

\newcommand{\ud}{\mathrm{d}} 

\newcommand{\showchg}{}

\usepackage{layouts}

\begin{document}
 

\title{\MyPaperTitle}


\author{Kristinn Torfason}

\author{Agust Valfells}

\author{Andrei Manolescu}
\affiliation{School of Science and Engineering, Reykjavik University, Menntavegur 1, IS-101 Reykjavik, Iceland}

\date{\today}

\begin{abstract}
High resolution molecular dynamics simulations with full Coulomb interactions of electrons are used to investigate field emission from a prolate spheroidal tip. The space charge limited current is several times lower than the current calculated with the Fowler-Nordheim formula. The image-charge is taken into account with a spherical approximation, which is good around the top of the tip, i.e. region where the current is generated.
\end{abstract}


\maketitle

\section{Introduction\label{sec:intro}}
Although the underlying physics of field emission have been understood for nearly 90 years~\cite{Fowler173} it is still a vibrant area of research,
particularly as field emitters have become important electron sources in modern devices~\cite{Zhu1.124541, Xu200547, 88525, 4804811}.
As field emission is highly dependent on the electric field at the cathode surface it can be strongly affected by the space charge from the emitted current.
Hence, the current density will lie somewhere between the values predicted by the Fowler-Nordheim equation~\cite{Forbes1.2827505}, where space charge effects are neglected,
and that predicted by the Child-Langmuir equation~\cite{PhysRevSeriesI.32.492, PhysRev.2.450, PhysRevLett.110.265007} where space charge reduces the surface electric field.
Considerable work has been done on the influence of space charge on field
emission~\cite{PhysRev.92.45, Lau1.870603, Forbes1.2996005, Rokhlenko1.3272690, Feng1.2226977, Jensen1.3692577, Torfason1.4914855, Jensen1.4921186, Rokhlenko1.4792059, Rokhlenko1.4847957, ZhuAng052106}.
The analysis in those papers has been almost exclusively based on one-dimensional models, although field emission typically takes place from some sort of protrusion due to local field enhancement.
It has been shown that for field emitter arrays, the one dimensional approximation is valid in the region over the array, above an elevation which corresponds to the spacing between emitters~\cite{Rokhlenko1.3272690, Jensen1.4921186}.
Closer to the emitter three dimensional effects must be taken into account. \citeauthor{ZhuAng052106}~\cite{ZhuAng052106} developed a self consistent model for continuous emission from a prolate spheroidal,
while \citeauthor{Jensen1.4921186}.~\cite{Jensen1.4921186} have studied discrete emission of rings from a hemispheric emitter.

In this paper we build on our previous work~\cite{Torfason1.4914855} to show how molecular dynamics (MD) based codes can be used to simulate field emission of electrons from a sharp tip. 
Although the MD approach is in principle computationally costly, it has the advantage of being able to account for the discrete emission of single electrons and Coulomb interaction between them.
These are likely to be important effects in the length scale describing a field emitter and its immediate environment.

The paper is organized as follows. \autoref{sec:method} of this paper gives a description of the model and simulation methodology used.
Simulation results are presented in~\autoref{sec:results} with a short summary and discussion in~\autoref{sec:summary}.
 
\section{Methodology\label{sec:method}}
\begin{figure}[b]
    \centering
    \includegraphics[]{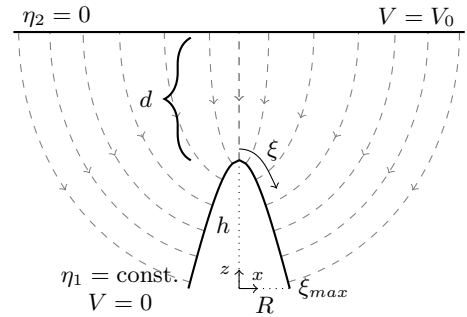}
    \caption{A sketch of the prolate spheroidal tip parameters and coordinate system.
    In this system a constant \(\eta\) defines a hyperbolic surface and \(\xi\) is a coordinate along that surface.
    To define the tips surface the parameters \(R\), \(h\) and \(d\) are used.
    They represent the base radius, the height of the tip and the gap spacing respectively.
    The dashed lines show the electric field lines in the system.}
    \label{fig:system}
  \end{figure}

  A prolate spheroidal coordinate system,
  \begin{equation}\begin{split}
    x &= a \sqrt{\xi^2-1}\sqrt{1-\eta^2} \cos{\phi}\, ,\\
    y &= a \sqrt{\xi^2-1}\sqrt{1-\eta^2} \sin{\phi}\, ,\\
    z &= a \xi \eta\, ,
  \end{split}\end{equation}
  where \(\xi \in [1, \infty]\), \(\eta \in [-1, 1]\), \(\phi \in [0, 2\pi]\) and
    \begin{equation}\label{eq:tip-a}
    a = \sqrt{\frac{d^2R^2}{h^2 + 2dh} + d^2}\, ,
  \end{equation}
  is used to define a spheroidal tip as depicted in~\autoref{fig:system}.
  Here \(R\) is the base radius and \(h\) the height of the tip.
  The gap spacing \(d\) is measured from the peak of the tip to the anode which is
  given by \(\eta_2 = 0\) and held at potential \(V_0\). The cathode is defined by a surface of
  constant \({\eta_1 = -d/a}\) and is at zero potential.
  
  The MD approach is used to calculate electron motion and it is also the basis for
  the field emission algorithm. \showchg{The simulation is of high resolution in the sense that every
  single electron present in the vacuum gap is treated as an individual particle,
  and the force acting upon it is calculated directly from the vacuum electric field and superimposed
  field stemming from every other free electron in the gap and image charge partners.
  Thus the total field at any point is \(E = E_{SC} + E_0\), where \(E_{SC}\) is the detailed space charge field which is
  a superposition of the electric field of the electric fields of all free electrons in the gap and image charge partners,
  and \(E_0\) is the vacuum electric field, which has a closed form analytic solution~\cite{pan:2151}. There is no spatial meshing involved.}
  Particle advancement is calculated using Verlet integration with a time step of \(1.0\,\mathrm{fs}\).

  Field emission is a quantum mechanical tunneling process and the resulting current density \(J\) can be described with the Fowler-Nordheim   
  equation~\cite{Fowler173}
  \begin{equation}\label{eq:FN-eq}
    J = \frac{A}{t^2(\ell)\phi}F^2 \mathrm{e}^{-\nu(\ell)B\phi^{\frac{3}{2}}/F}\, ,
  \end{equation}
  where \(\phi\) is the work-function and \(F\) is the field
  at the surface of the cathode, taken to be positive. \(A = e^2/(16\pi^2\hbar)\;
  [\mathrm{A}\,\mathrm{eV}\,\mathrm{V}^{-2}]\)
  and \(B = 4/(3\hbar) \sqrt{2m_e e}\;
  [\mathrm{eV}^{-\frac{3}{2}}\,\mathrm{V}\,\mathrm{m}^{-1}]\) are the
  first and second Fowler-Nordheim constants, while \(\nu(\ell)\)
  is called the Nordheim function and arises due to the
  image-charge effect. It contains complete elliptic integrals of
  the first and second kind and is related to \(t(\ell)\) by the
  relation \(t(\ell) = \nu(\ell) - (4/3)\ell\, \ud \nu(\ell) / \ud
  \ell\). We use the approximations found by~\citeauthor{Forbes08112007}~\cite{Forbes08112007},
  \begin{subequations}\label{eq:nordheim-fun}
  \begin{equation}
  \nu(\ell) = 1 - \ell + \frac{1}{6} \ell \ln(\ell)
  \end{equation}
  and
  \begin{equation}
  t(\ell) = 1 + \ell\left(\frac{1}{9} - \frac{1}{18}\ln(\ell) \right) ,
  \end{equation}
  where
  \begin{equation}\label{eq:ell}
    \ell = \frac{e}{4\pi\varepsilon_0}\frac{F}{\phi^2}\,.
  \end{equation}
  \end{subequations}
  
  Image-charge is taken into consideration in our MD simulations with two objectives in mind.
  First, the image-charge of the electron being emitted is taken into account by
  calculating the barrier potential which the electron must tunnel through by use
  of~\autoref{eq:nordheim-fun}.
  \begin{figure}[!ht]
    \centering
    \includegraphics[]{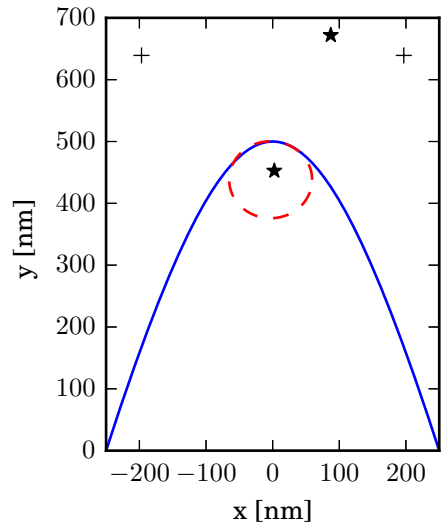}
    \caption{The spherical image-charge approximation. The \textcolor{blue}{blue solid}
    line shows cross section of the tip. While the \textcolor{red}{red dashed} line
    shows the imaginary sphere used to calculated the image-charge partner for the \(\star\).
    The \(\star\) shows the location of the electron and its image-charge partner used
    in~\hyperref[fig:ic-sphere-theory]{\autoref{fig:ic-sphere-theory}(a)},
    while the +'s show the positions of the electrons used
    in~\hyperref[fig:ic-sphere-theory]{\autoref{fig:ic-sphere-theory}(b)}.}
    \label{fig:sphere-image-charge-ex}
  \end{figure}
  Second, to maintain the proper boundary conditions at the cathode, we include
  a spherical image-charge approximation for the space-charge to be found in the gap.
  In this approximation each electron is given an image-charge partner whose  position is
  calculated using the image-charge equations for a sphere. The radius of curvature of the tip surface at
  the point on the tip closest to the electron is calculated. The sphere is then placed inside the tip, tangent
  at this point, its radius being equal to the radius of curvature at the contact point.
  An examples of this method can be seen in~\autoref{fig:sphere-image-charge-ex}.
\begin{figure}
  \centering
  \includegraphics[]{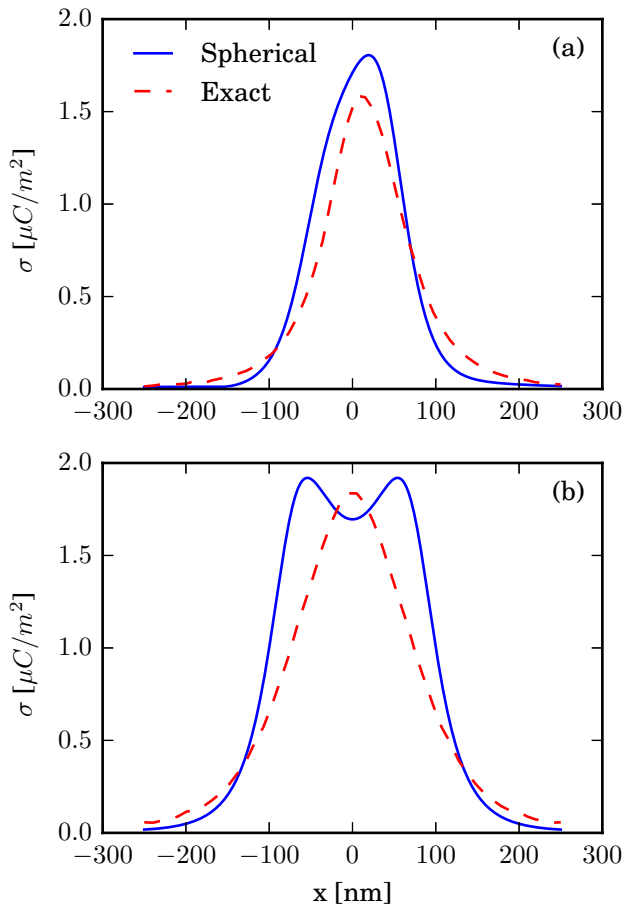}
  \caption{Comparison of the surface charge density on the surface of the tip for the 
           spherical image-charge approximation (\textcolor{blue}{blue solid}) and
           exact solution~\cite{Peridier4888, Gil2012794} (\textcolor{red}{red dashed}).
           (a) Electron at (\(\xi = 1.01\), \(\eta = 0.8\)) close to the peak of the tip
           (Shown as a \(\star\) in~\autoref{fig:sphere-image-charge-ex}).
           (b) Two electrons at (\(\xi = 1.05\), \(\eta = 0.8\)
           (Shown as the left and right \(+\) in~\autoref{fig:sphere-image-charge-ex}).
           The surface charge density at the tip of the peak due to the vacuum field is
           \(8.2\times 10^4\, \mathrm{\mu C/m^2}\).}
  \label{fig:ic-sphere-theory}
\end{figure}
\begin{figure}
  \centering
  \includegraphics[]{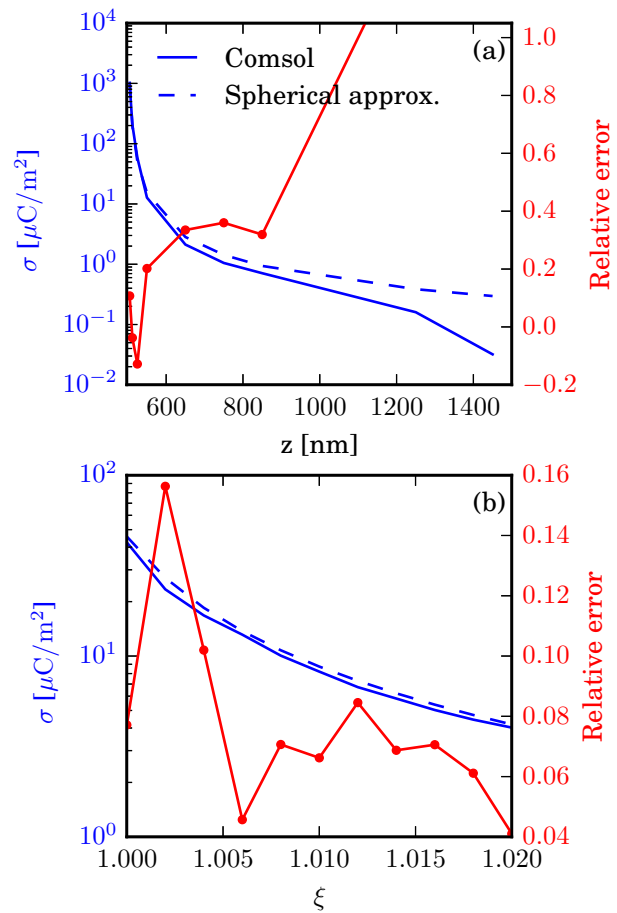}
  \caption{The relative error (\textcolor{red}{right axis}) of the surface charge density (\textcolor{blue}{left axis})
  at the peak of the tip.
  a) Varying \(z\) with \(\xi = 1\) fixed. b) Varying \(\xi\) with \(\eta = -0.95\) fixed.}
  \label{fig:rel_err}
\end{figure}
 \begin{figure}
    \centering
    \includegraphics[]{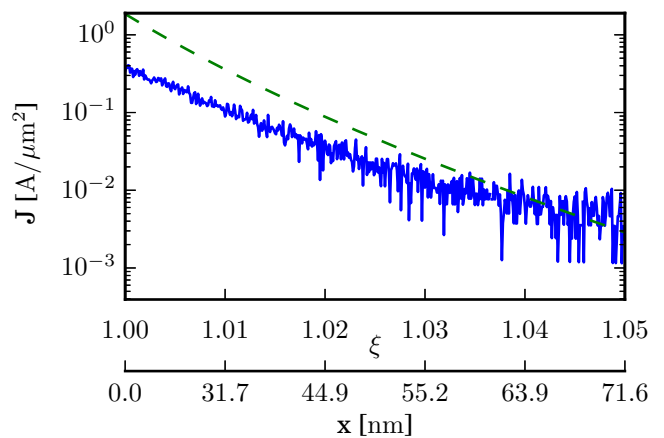}
    \caption{Current density as a function of \(\xi\), the position on the tip
    with the distance from the peak shown for comparison. The \textcolor{blue}{blue}
    line shows the simulations results, while the \textcolor{green}{green} dashed line
    shows the Fowler-Nordheim result of~\autoref{eq:FN-eq}.}
    \label{fig:J-tip}
  \end{figure}  
An exact open-form solution~\cite{Peridier4888, Gil2012794} of the image-charge problem
  for the prolate spheroidal tip exists. However, in practice it is not usable for MD simulations since its numerical convergence is
  too slow.
  In~\autoref{fig:ic-sphere-theory} we compare the
  exact results for the surface charge density to our spherical image-charge approximation. As can be seen the approximation
  works best near the peak of the tip and becomes less effective when electrons move further
  away from the peak as in~\hyperref[fig:ic-sphere-theory]{\autoref{fig:ic-sphere-theory}(b)}.
  In spite of the discrepancy this approximation works out quite well, since most of the emission takes place near the peak
 (to be discussed later, see ~\autoref{fig:J-tip}) and the electrons closest to the peak contribute the most to the field there.
  \autoref{fig:rel_err} shows the relative error of the surface charge density at the peak of the tip.
  In~\hyperref[fig:rel_err]{\autoref{fig:rel_err}(a)} the electron is placed at \(\xi = 1.0\) and
  the charge density is calculated for various heights above the tip.
  The exact solution had troubles converging for some locations of the electron and therefor the tip was also
  modeled in COMSOL~\footnote{COMSOL Multiphysics \url{https://www.comsol.com}} to find the charge density.
  \showchg{COMSOL solves the Laplace equation using finite element analysis with a grounded conducting boundary at the tip,
  and therefore can be used to calculate the same surface charge density as the the image charge method.
  By comparison of the two methods for calculating the surface charge density,
  we can see from~\hyperref[fig:rel_err]{\autoref{fig:rel_err}(a)} that the relative error
  is small when the electron is close to the tip and increases as the electron moves away.} Since the influence of
  the electron is strongest when it is close to the peak, the large relative error of the spherical model at locations far away from the tip
  have negligible effect on the simulation results. \hyperref[fig:rel_err]{\autoref{fig:rel_err}(b)} shows the relative error of the
  charge density at the peak with \(\eta = -0.95\) fixed and \(\xi\) is varied from \(1.000\) to \(1.020\). The electron
  is located quite close to the tip and we see that the relative error is small even if the electron
  moves further down towards the tip surface. 
  \showchg{The reader should note that the errors shown in~\autoref{fig:ic-sphere-theory} and~\autoref{fig:rel_err} are not due to numerical effects,
  such as meshing or step size, but because of the adoption of the spherical conductor model for the image charge.
  To elaborate, when the electron is far down along the side of the tip and placed at a distance from it on the order of the base radius of the tip,
  the image charge will be placed near the other side of the tip, or even outside the tip. This can lead to a large error in the image charge calculation,
  but due to the fact that a negligible amount of electrons are to be found in that position, the effect of this error on the overall simulation is small.
  A comparison of~\autoref{fig:sphere-image-charge-ex} and~\autoref{fig:J-tip} will show the reader how field emission is tightly bound to the apex of the tip.}
  
%
\section{Results\label{sec:results}}
\begin{figure}
    \centering
    \includegraphics[]{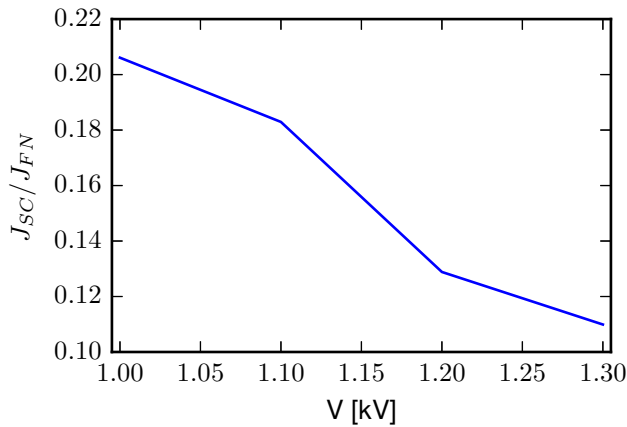}
    \caption{Current density versus voltage at the peak of the tip. The current density, from the field including space charge, is scaled with the Fowler-Nordheim current density at the peak calculated from the vacuum field.}
    \label{fig:JvsV}
  \end{figure}

  We examine a system with the parameters \(R = 250\,\mathrm{nm}\), \(h = 500\,\mathrm{nm}\),
  \(d = 1000\,\mathrm{nm}\), \(\phi = 4.7\,\mathrm{eV}\) and \(V_0 = 1\,\mathrm{kV}\), \showchg{which gives
  a surface charge density of \(8.2\times 10^4\, \mathrm{\mu C/m^2}\) at the peak.}
  Care was taken in selecting the vacuum field such that the parameter \(\ell\) in~\autoref{eq:ell}
  was less than \(1\). If \(\ell\) is larger than \(1\) then
  the tunneling barrier will be below the Fermi energy. In~\autoref{fig:J-tip} the current density is plotted
  as a function of \(\xi\), the position on the tip. The results of the simulations are given by the \textcolor{blue}{blue}
  solid line.
  We see that the emission is largest at the peak of the tip (\(\xi = 1\)) and falls off rapidly. This was to be expected
  as the field is strongest at the peak. At around the value of \(\xi = 1.02\) the current density
  has dropped by about 95\% of the peak value.
  The results obtained are similar to the results from~\citeauthor{ZhuAng052106}~\cite{ZhuAng052106}
  where they use a fluid like model. The \textcolor{green}{green} dashed lines shows~\autoref{eq:FN-eq} plotted
  for the Vacuum field. We can see that space charge limited current density is about 20\% of the current density
  calculated from the Vacuum field. 
  This can also be seen in~\autoref{fig:JvsV}, where the relationship between the scaled current density and voltage is shown. The current density has
  been scaled with Fowler-Nordheim current density for the vacuum field. It is clear from the figure that as the voltage is increased space charge effects become stronger.

\begin{figure}
    \centering
    \includegraphics[]{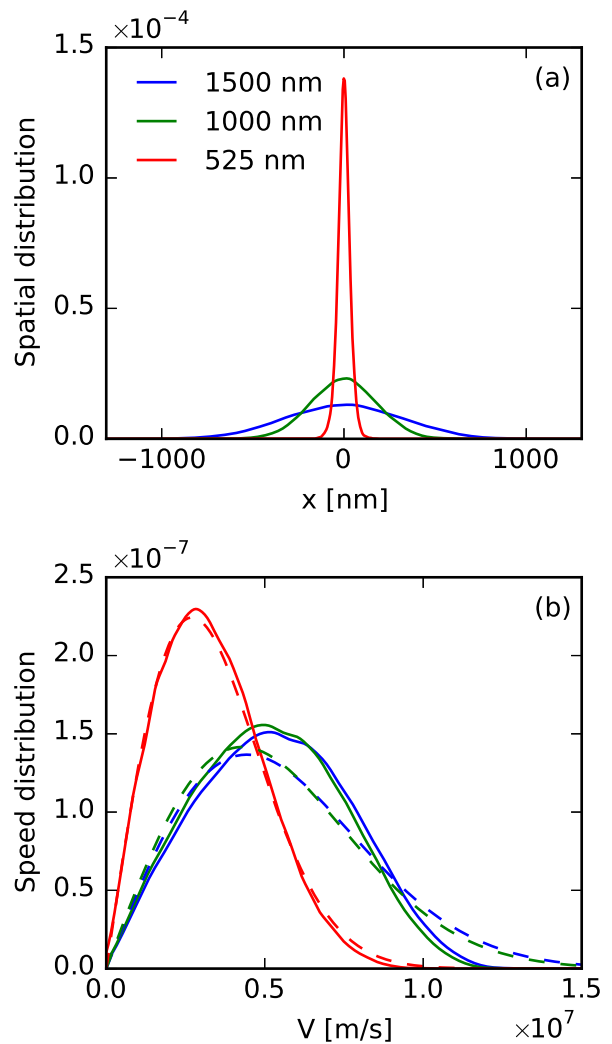}
    \caption{The cross-section of the spatial and the transverse speed distribution. Taken through planes
    at three different heights \(z = 525\), \(1000\) and \(1500\,\mathrm{nm}\) over the duration of the simulation.
    (a) The spatial distribution of electrons.
    (b) The transverse speed distribution of electrons, 
    with solid lines the MD simulations and dashed lines fitted Maxwell-Boltzmann distributions.}
    \label{fig:spatial-speed-dist}
  \end{figure}

  In~\hyperref[fig:spatial-speed-dist]{\autoref{fig:spatial-speed-dist}(a)} we see a cross-section of the spatial distribution
  of electrons taken at three different elevations, \(z = 525\), \(1000\), and \(1500\,\mathrm{nm}\).
  The distributions are normalized such that the area  under the curve is 1.
  The figure shows that the electrons spread further apart the higher they are. This spread comes from the Coulomb
  repulsion between electrons and the transverse component of the vacuum field.
  \hyperref[fig:spatial-speed-dist]{\autoref{fig:spatial-speed-dist}(b)} shows the transverse speed distribution
  for the same three elevations. The solid lines are the simulations results, while the dashed lines are curves fitted to
  the 2D Maxwell-Boltzmann distribution. We see that closest to the peak of the tip the distribution fits quite well with
  the Maxwell-Boltzmann distribution. Electrons are emitted from the tip with zero velocity and it seems they fit to the
  Maxwell-Boltzmann distribution quite quickly. At higher elevations, after the electrons have been accelerated longitudinally by the applied field, the transverse speed distribution is no longer described by the Maxwell-Boltzmann distribution.
  The temperatures obtained from the fitting of the curves are \(T_{1500} = 1.3\times10^6\,\mathrm{K}\),
  \(T_{1000} = 1.2\times10^6\,\mathrm{K}\) and \(T_{525} = 0.48\times10^6\,\mathrm{K}\).

\begin{figure}
    \centering
    \includegraphics[]{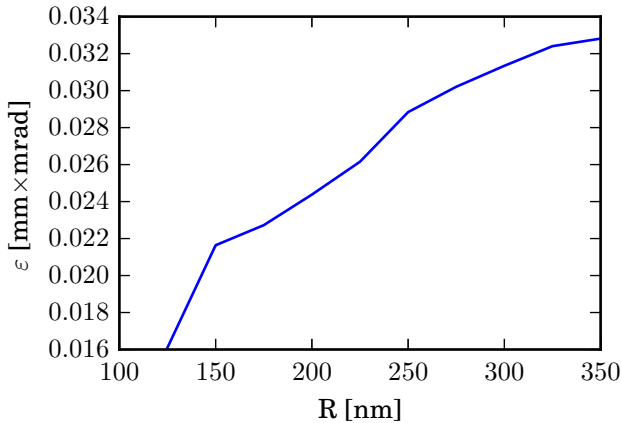}
    \caption{The emittance in the system as a function of the base radius. The electric field at the peak is the same
             in all simulations.}
    \label{fig:emittance}
  \end{figure}

  The MD simulations also allows us to calculate the emittance of the beam coming from the tip.
  The RMS emittance~\cite{MartinBeamOptics} is calculated using \(\epsilon_x = \sqrt{\overline{x^2}\;\overline{x^{\prime 2}} - \overline{xx^{\prime}}^2}\),
  where \(x\) is the position and \(x^\prime = \ud x / \ud z \approx P_x / P\) is the slope of the trajectory. Due to the symmetry in the system
  calculating \(\epsilon_y\) for the \(y\)-direction gives almost the same results as for the \(x\)-direction. The emittance \(\epsilon\) is given
  as the average of \(\epsilon_x\) and \(\epsilon_y\).
  The averages of \(\overline{x^2}\), \(\overline{x^{\prime 2}}\) and \(\overline{xx^{\prime}}^2\) are taken over the duration of the simulation.
  In~\autoref{fig:emittance} the emittance is shown as a function of the base radius of the tip.
  The voltage is also varied in such a way to keep the same electric field at the peak for all runs.
  The emittance grows as the base radius increases. Since the electric field at the peak is the same for all runs the active emission
  area on the tip becomes larger with increasing radius. More electrons are then emitted further away from the peak which increases
  the spread of the beam and the emittance.

\section{Summary and conclusions\label{sec:summary}}
The MD approach to simulation offers a high fidelity model of field emission from a tip and charge propagation through the gap of the diode.
Discrete particle effects, such as single electron emission and Rutherford scattering are implicitly included.
The drawback to the MD approach is the computational  cost involved, which scales as the number of particles squared.  
In the case of the prolate hyperbolic spheroid tip, the issue of computational cost is greatly exacerbated
when the image charge for electrons in proximity to the tip is included by using exact open form equations for the image charge.
To get around this problem, we have made the approximation that the image charge of an electron is calculated from an osculating sphere,
tangent to the tip surface at the point closest to the electron. This speeds up computation significantly, and makes the MD approach practical.
Although this approximation is quite good for electrons near the tip,
it becomes bad for electrons that are placed to the side of the tip and can even lead to the image charge being situated outside the tip.
Nevertheless this is not of great concern since almost all of the emission takes place from the point of the tip. Hence, errors due to spurious electrons emitted to the side are minimal.
\showchg{In the region near the apex of the tip, the relative error in calculating the space charged induced surface charge density using the spherical approximation ranges from 4 --- 15\%.
For practical considerations one may surmise that this is not a large error compared to errors due to the fact that field emitters are not exact hyperbolic spheroids, the work function may vary etc.
The approximation should be good enough to give users an adequate tool to examine the physics of discrete particle effects in the proximity of an emitting tip.}

The utility of the MD approach is that its advantageous inclusion of discrete particle effects dovetails nicely with the limitations
set by computational cost. This is due to the fact that discrete particle effects are mostly important within a region on the order of a Debye length,
that typically includes a number of particles that is manageable for MD calculations.
We can comfortably handle several thousands of electrons on an average high performance computer cluster, which is a realistic number for a nanometric diode.
Larger systems with more particles can be accurately approximated with continuous models or by using particle-in-cell simulations (PIC).
For the tip geometry described in this work, a motivating factor is that MD simulations may be used to
describe three dimensional field emission in the immediate vicinity of a single emitting tip in an array,
and those simulations used to describe the first cell in a larger PIC code for modeling of field emitter
arrays in a manner similar to that described by~\citeauthor{Jensen1.4921186}~\cite{Jensen1.4921186}.
The length scale for such a cell is that of the distance between neighboring emitters, and as such is quite amenable to MD simulations.

\begin{acknowledgments}
  This work was financially supported by the Icelandic Research Fund grant number 120009021.
  The simulations were performed on resources provided by the Nordic High Performance Computing (NHPC).
\end{acknowledgments}

\bibliography{Vacuum-FE-Tip}
\end{document}